\documentclass[reprint, amsmath, amssymb, aps]{revtex4-1}
\usepackage{graphicx}
\usepackage{dcolumn}
\usepackage{bm}
\usepackage{bbold}
\usepackage{slashed}
\usepackage{soul}
\usepackage{times}
\usepackage{epstopdf}
\usepackage{epsfig}
\usepackage{graphicx}
\graphicspath{ {images/} }
\usepackage{amsmath}
\usepackage[shortlabels]{enumitem}
\usepackage[titletoc, title]{appendix}
\usepackage[colorlinks,
            linkcolor=red,
            anchorcolor=blue,
            citecolor=green
            ]{hyperref}

\begin{document}
\title{Parametrically narrow tetraquark states containing one heavy quark and one heavy antiquark}
\author{Yiming Cai}
\email{yiming@umd.edu}
\author{Thomas Cohen}
\email{cohen@physics.umd.edu}
\affiliation{Maryland Center for Fundamental Physics and the Department of Physics,\\
University of Maryland, College Park, MD, USA\\}

\date{\today}

\begin{abstract}
This paper focuses on tetraquarks containing one heavy quark and one heavy antiquark in the formal limit where the heavy quark masses go to infinity. It extends the theoretical framework developed for tetraquark containing two heavy quarks based on semiclassical analysis and the Born-Oppenheimer approximation.  It is shown that for sufficiently small light quark masses, parametrically narrow $q\bar{q}'Q\bar{Q}$ tetraquarks must exist in the formal limit of arbitrarily large heavy quark masses.  The analysis is model-independent; it requires no assumptions about the structure of the state or the mechanism of its decay beyond what can be deduced directly from QCD.   The formal analysis here is only directly applicable to tetraquarks with large angular momentum and hence does not directly apply to experimentally observed tetraquark candidates.  We also discuss the condition needed for a version of the analysis to apply for low angular momenta states.  Whether this condition holds can be explored via lattice calculations.  Current lattice calculations suggest that the condition is not met.  If these preliminary calculations are confirmed, there are strong implications for the nature of the observed tetraquark resonances containing $\overline{c}c$ or $\overline{b}b$ pairs. It raises the possibility that the observed tetraquarks exist and are narrow not because the charm and bottom quarks are extremely heavy, but because they are sufficiently light. \end{abstract}

\maketitle

\section{Introduction}
Understanding exotic hadrons is an important branch of the current hadron structure and spectroscopy research.  This paper focuses on exotic hadrons containing one heavy quark $Q$ and one heavy antiquark $\Bar{Q}$ in the formal limit where the heavy quark masses are taken to be arbitrarily large. The goal is to demonstrate that, when the light quark masses are sufficiently light, $q\Bar{q}'Q\Bar{Q}$ tetraquark states must exist as parametrically narrow resonances in the heavy quark mass limit.

Since the discovery of the exotic hadron X(3872) by Belle in 2003 \cite{choi2008sk}, numerous other exotic candidates have been
found in the heavy quarkonium sector\cite{ali2017exotics}, which are named XYZ states such as  $Z(3900)$\cite{ablikim2013observation,liu2013study, xiao2013observation}, $Z(3885)$\cite{ablikim2014m, ablikim2015confirmation, ablikim2015observation}, $Z(4020)$\cite{Z4020ablikim2013observation, Z4020ablikim2014m} and  $Z(4430)$\cite{choi2008observation,chilikin2013experimental, aaij2014observation,mizuk2009r}. However, even for $X(3872)$, the most thoroughly studied of this class, its physical nature is still controversial. Because of its remarkable closeness to the $D^0\Bar{D}^{*0}$ threshold, there has been significant  effort to describe the $X(3872)$ as primarily $D^0 \Bar{D}^{*0}$ molecular\cite{guo2018hadronic}, though arguments exist for questioning this interpretation and suggesting a large heavy quarkonium component\cite{lebed2017heavy}. The large amount of experimental  tetraquark signal containing $c\Bar{c}$ or $b\Bar{b}$ components makes it an important task to theoretically understand a tetraquark of $q\Bar{q}'Q\Bar{Q}$ type.

On the theoretical side, though quantum
chromodynamics (QCD) has long been accepted as the underlying theory for hadron physics, due to its nonperturbative nature, QCD cannot currently be used to determine the properties of tetraquark states. Nonperturbative numerical lattice QCD simulations have  difficulties dealing with such heavy tetraquark state, and the current state-of-the-art lattice QCD simulations with the pion mass $m_{\pi} = 266$ MeV are still unable to settle down the existence and structure of $X(3872)$ state\cite{padmanath2015x}. 

In the absence of direct QCD calculations, one might hope to learn about these states via models that attempt to capture the essential physics of QCD in a tractable form---for example as a few-body quantum mechanical system.  While there is much of interest that can be learned from such models, they suffer from the generic difficulty that the results are highly dependent on the potentially unreliable assumptions that are put into the model at the outset.

A successful model needs to compute more than simply the mass of the resonance; it needs to be able to describe its width, {\it i.e.}  accurately model the decay of the state.  Often hadronic models treat the problem of the structure of the state and its decay as two essentially distinct problems---the structure of the state and its mass is computed while neglecting the possible decay and then a separate model for the decay is used to compute the width perturbatively. Model builders may have more intuition about the structure of the state than about the detailed decay mechanism.   There is a basic intuitive sense that the reason that tetraquark states are narrow enough to discern is that the decay channels ({\it e.g.}  a quarkonium state plus a pion) are qualitatively very different from the dominant structure composing the state ({\it i.e.} a molecule state of D mesons) making the coupling between the final state and the basic structure small.  However, this intuition is very difficult to quantify or model in a reliable way.  One difficulty is that final states will typically contain pions which {\it a priori} suggests that appropriate modeling requires a correct treatment of chiral symmetry, while typical models of the underlying structure of the tetraquark state are few-body models based on quantum mechanical wave functions, for which it is very difficult to encode the physics of chiral symmetry correctly.

Given these difficulties, it is reasonable to seek out regimes in QCD in which the question of a tetraquark's existence as a narrow resonance is tractable from first principles.  Finding such regimes is useful even if these do not enable one to compute energies and widths directly in that regime and even if the regimes do not correspond closely to the physically observed states.  They may still serve to give some insight into why the physically observed states exist.     Since one of the defining characteristics of the recent tetraquark candidates is that they contain a heavy quark and a heavy antiquark, one reasonable conjecture is that the reason tetraquark states of this sort exist as narrow resonances is tied directly to the large mass of the heavy quark.  This paper is intended to give some insight into this question by showing in a model-independent way that in the formal heavy quark limit a certain class of tetraquarks containing a heavy quark and antiquark must exist as parametrically narrow resonances.  However, at the same time, we observe that this model-independent analysis is only valid for resonances with angular momenta that are parametrically large.

Model-independent analyses demonstrating the existence in the heavy quark mass limit $m_Q \rightarrow \infty$ of tetraquarks with configuration $\Bar{q}\Bar{q}'QQ$ have long been known \cite{cohen2006doubly, manohar1993exotic, zouzou1986four}. The lowest such states are formally stable against strong interaction.  Recently it was observed that for very large heavy quark masses, there are numerous such tetraquark states with given quantum numbers and the higher-lying ones of these are not stable against strong decays\cite{cai2019existence}---they can decay via pion emission into lighter tetraquarks.   Such unstable states were shown to be parametrically narrow, the resonances become increasingly discernible as the heavy quark limit is approached.   Of course, due to the lack of experimental evidence, $\Bar{q}\Bar{q}'QQ$-type tetraquarks are currently of less phenomenological interest than the  $q\Bar{q}'Q\Bar{Q}$-type.  Moreover, given realistic quark masses, it seems unlikely that multiple $\Bar{q}\Bar{q}'QQ$-type tetraquarks with fixed quantum numbers exist in nature.  Nevertheless the analysis may be of some interest as it shows that,  at least in some regime---albeit an unphysical one---of QCD, unstable tetraquarks exist as resonances with widths can be seen to be parametrically narrow simply due to the large mass of the heavy quark.  
 
The purpose of the present paper is to show that a somewhat similar sort of analysis can be applied to a class of tetraquarks of the $Q\Bar{Q}$ type.  In particular, it is shown in a model independent way that exotic tetraquarks of the $Q\Bar{Q}$ type must exist as parametrically narrow resonances in the formal heavy quark limit---at least if the pion is sufficiently light.  The argument, as formulated, applies to tetraquarks with angular momentum in a prescribed range; the angular momentum needs to be large---it scales with $m_Q^{1/2}$---but cannot exceed a critical value (which also scales with  $m_Q^{1/2}$).   The argument is model-independent in the sense that it does not depend on the details of the system, such as an explicit model of the short-distance interaction or a detailed model of the mechanism of decay.  Rather, this work is a QCD argument, which depends on only well-established features of the theory including the validity of chiral perturbation theory to describe the longest-range phenomena when the light quark masses are sufficiently light, and heavy quark effective theory when the heavy quark masses are sufficiently large.  The analysis makes use of semiclassical analysis and the Born-Oppenheimer approximation, which can be shown to be valid in the formal regime we study.  

The type of tetraquark that emerges in this analysis is special.  The dynamics are dominated by the motion of the heavy quarks which emerge as the effective degrees of freedom; as the heavy quarks move they drag the light quarks with them in a fixed coherent way.  Several types of models of tetraquark are consistent with this, for example, what model builders often refer to as ``molecules''--states which can largely be described as bound states of heavy mesons or ``diquark'' descriptions in which the relative position of quasiparticles containing a heavy and light quark (or a heavy and light antiquark) is the dominant degree of freedom.  Note that as a matter of principle, molecular-type and diquark-type tetraquarks in general cannot be distinguished from other types since they have the same quantum numbers.  However, as the heavy quark limit is approached the distinction between a class of dynamics that includes both molecular-type and diquark-type tetraquarks and other possible types becomes increasingly sharp.

\begin{figure*}
\centering
\includegraphics[width=.6 \textwidth]{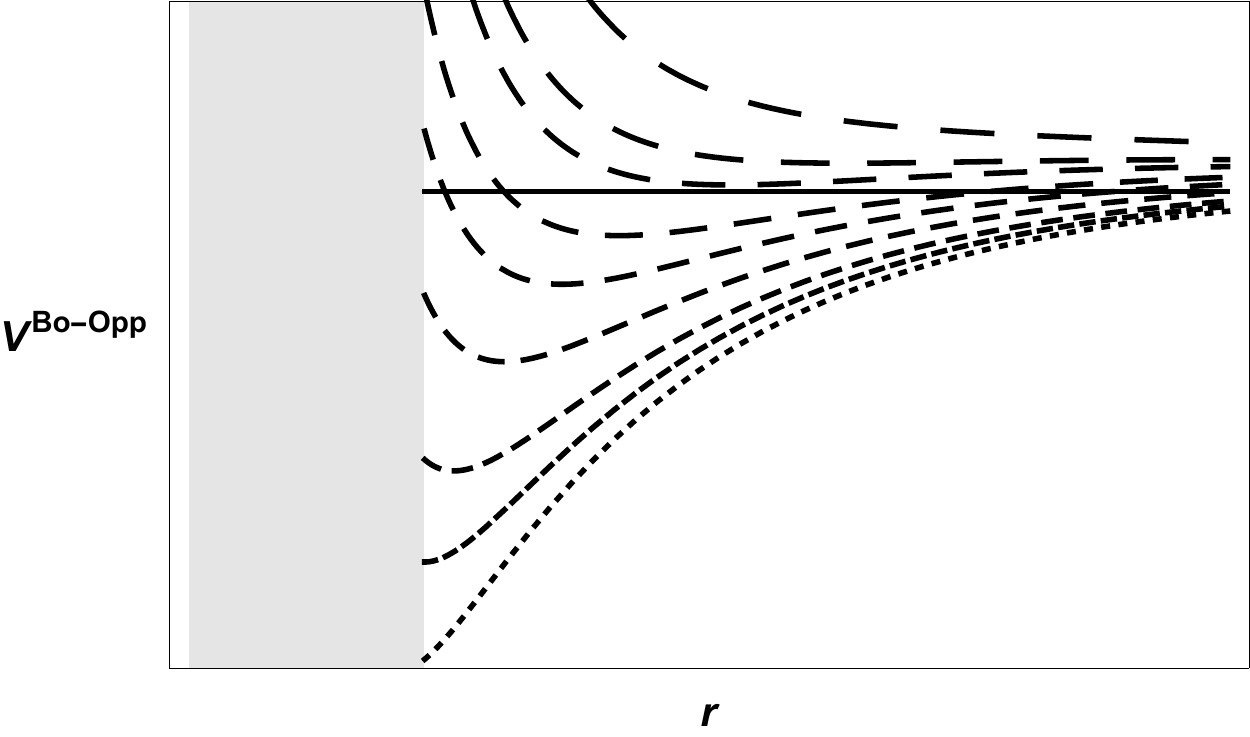}
    \caption{A schematic representation of the Born-Oppenheimer effective potential for various values of the angular momentum of order $\sqrt{m_Q/\Lambda}$, where $\Lambda$ is a typical hadronic scale.  Longer dashes correspond to higher  angular momenta.  The gray shaded region corresponds the gapless domain where it is energetically favorable to emit a pion.  The underlying potential is not realistic and is to chosen for the purpose of illustrating qualitative behavior.  The horizontal line corresponds to the value at infinity}
    \label{cartoon}
\end{figure*}

A quick preview of how this result comes about: consider a system with a heavy quark and antiquark which has isospin one; by quantum numbers it must contain light quarks.  When the heavy quarks are sufficiently massive, they move slowly compared to the other degrees of freedom in the problem.  Provided that the spectrum for the remaining QCD degrees of freedom at a fixed separation of the heavy quarks has a gap between the ground state for fixed quantum numbers and the first excited state, a Born-Oppenheimer type of analysis will be a valid description of QCD with the separation between the heavy quarks as the degree of freedom.

 At a sufficiently large separation, a gap is guaranteed to exist on physical grounds: as the separation gets large the system behaves as two well-separated heavy mesons.  At small enough separation the system is guaranteed to be gapless since it will physically behave as a continuum pion scattering state against the closely spaced heavy quarks-antiquark pair as one would have an onium state.  Thus, the Born-Oppenheimer potential is defined beyond a certain value of separation and not below it.   If there exists a state whose typical separation is dominantly in the region where the Born-Oppenheimer potential exists, its distribution is well localized and its time evolution keeps the state in that localized region for a long time,  then narrow tetraquarks exist. Standard semiclassical implies that if the effective Born-Oppenheimer potential has a local minimum at some fixed separation in the region where it is well defined then as the heavy quark mass gets large,  multiple narrow tetraquarks of this type must exist; such states having the heavy quarks separation are well localized near the minimum of the potential.   

 The form of this effective potential Born-Oppenheimer potential depends on the angular momentum of the system due to a centrifugal barrier.    One can show generically that if there is an attractive region of the potential without the centrifugal barrier---something that is guaranteed to happen in certain channels for sufficiently small pion masses---then between a minimum and a maximum value of angular momentum order $\sqrt{m_Q/\Lambda}$ (where $\Lambda$ is a typical hadronic scale), a local minimum in the Born-Oppenheimer potential will develop implying narrow tetraquarks. This situation is illustrated schematically in Fig.~\ref{cartoon}.

There are two generic features concerning the parametric dependence on the mass of the heavy quarks that naturally arise for tetraquarks that emerge from this type of analysis.  The first feature is that the width of the tetraquark resonance decreases with the mass of the heavy quark and becomes arbitrarily narrow as the mass goes to infinity.  This is intriguing in light of the fact that widths of observed tetraquarks appear to be narrow.  The second feature is that the number of distinct tetraquarks with fixed quantum numbers grows parametrically with the mass of the heavy quark and diverges in the heavy quark limit.    

Though the artificial limit where the heavy quark mass is arbitrarily large is not well satisfied in the real world, and tetraquarks with large angular momenta are not directly related to experimental tetraquark candidates, this result may give some insight into the physically relevant important tetraquark configuration $q\Bar{q}'Q\Bar{Q}$. This could be viewed in the same spirit as large $N_c$ QCD. Though in the physical world $N_c$ is not large enough for all behaviors in the large $N_c$ limit to have identifiable real-world analogs, it still serves as an important theoretical tool to understand the rich structure of QCD. 

The restriction to a limited range of large angular momentum in the present analysis is unfortunate, as it makes the analysis inapplicable to the case of low angular momenta as seen in the observed exotics containing $c \overline{c}$  or  $b \overline{b}$ pairs.  We consider a condition by which this analysis may be extended to the small momentum case and argue that if the condition is not met, the analysis may suggest that narrow tetraquarks containing an arbitrarily heavy quark and antiquark of the type considered here do not exist.  We note that whether the condition holds can be explored via lattice simulations.  The current state-of-the-art lattice simulations\cite{Prelovsek:2019ywc} suggest that the condition may well not hold.  If a more complete lattice study shows compellingly that is the case, then we will have learned something significant about the observed heavy tetraquarks.  It suggests that the conjecture that the existence of these heavy tetraquarks as a narrow resonance as being tied to the charm and bottom quarks being extremely heavy may need to be turned on its head: their existence could be tied to the charm and bottom quarks being sufficiently light.

This paper begins with a brief review of the theoretical framework in Ref.\cite{cai2019existence}, which sets the stage for the analysis in this paper. The next section discusses conditions that allow one to extend the previous work on $\Bar{q}\Bar{q}'QQ$ to $q\Bar{q}'Q\Bar{Q}$ tetraquark in a simple way.  This will occur if the Born-Oppenheimer-type wave function that emerges from the analysis is effectively confined to a region in space in which the two heavy quarks are sufficiently far from each other, so that rearrangement effects that violate underlying assumptions of the analysis are suppressed.      Following this a key argument will be shown: for sufficiently large heavy quark masses that allow a Born-Oppenheimer description at large distances and sufficiently small light quark masses, there always exists a range of angular momenta for which the Born-Oppenheimer effective potential in a channel with isospin one  $q\Bar{q}'Q\Bar{Q}$ tetraquark quantum numbers will have a local minimum with the quarks separation of order $m_\pi^{-1}$.   This distance is sufficiently large that wave functions localized near it---as happens with heavy quark masses---will be localized in a regime where the Born-Oppenheimer treatment is self-consistent and rearrangement effects associated with the decay of the tetraquark are formally suppressed. The conclusion is that narrow $q\Bar{q}'Q\Bar{Q}$ tetraquark states must exist in the heavy quark limit of QCD when the light quark masses are sufficiently light. The paper concludes with a discussion section that explores the extension of this analysis to the more phenomenologically interesting case of small angular momenta; we formulate the necessary condition for this, discuss how lattice simulations can help determine whether it is met, and discuss the implications if it is not.

\section{Review of the analysis for $\Bar{q}\Bar{q}'QQ$ tetraquarks }
It is useful to review the analysis of  Ref.\cite{cai2019existence}, which shows that, parametrically narrow tetraquark $\Bar{q}\Bar{q}'QQ$ that are parametrically close to threshold must exist in the heavy quark mass limit, as the present analysis is an extension of it.   Ref.\cite{cai2019existence} is based on Born-Oppenheimer and semiclassical considerations, and the current paper adopts this framework. The original question of whether tetraquark states exist is a complicated quantum field theory problem; however, it can be shown that it can be reduced to an effective two-particle description formalism in the heavy quark limit, which will greatly simplify the discussion. 

Ref.\cite{cai2019existence} first {\it assumes} the system is in a regime in which the complicated quantum field theory problem can be simplified and powerful tools like semiclassical analysis and Born-Oppenheimer approximation can be used.  It is then shown that in this regime, multiple tetraquark states exist in the heavy quark limit including some parametrically close to the threshold which are above the threshold for decays into lighter tetraquarks plus pions.   The crux of the analysis is a demonstration that the parts of the field-theoretic Hilbert space that are neglected in passing to the Born-Oppenheimer description as a two-body quantum mechanics problem are suppressed, in the sense that their inclusion can only lead to decay rates of the unstable tetraquarks that are parametrically small in the inverse heavy quark mass.  This in turn shows that the original assumption was self-consistent.

The regime assumed (and subsequently shown to be consistent) in Ref.\cite{cai2019existence} is that the  $\Bar{q}\Bar{q}'QQ$ system requires that the heavy quarks be sufficiently massive that the creation of heavy quark-antiquark pairs can be neglected, leading to separate conservation of the number of heavy quarks and antiquarks.  When restricted to the sector of two heavy quarks, the quarks need to be heavy enough so that their motion is sufficiently slow to ensure that the light degrees of freedom can adjust adiabatically.  This reduces the problem to an effective quantum mechanical problem with a degree of freedom associated with the position of the heavy quark (which is arbitrarily well defined for states with tetraquark quantum numbers in the heavy quark limit of QCD).  In such a case, with fixed light quark quantum numbers, one can write an effective potential for a nonrelativistic quantum system with one degree of freedom associated with the separation $R$.  In such a Born-Oppenheimer-type approximation,  the tetraquark is described as a bound state of this effective potential.  Moreover, one knows that as $R \rightarrow \infty$ the system becomes two heavy mesons (the analog of the $D$ or $D^*$ which differ only due to the spin of the heavy quark which is irrelevant in the heavy quark limit).  

Thus, the effective potential at long-distance regime asymptotes to the long-distance interaction between two heavy mesons, which at the longest distances is known to be a one-pion-exchange potential.    As it happens, when the heavy quark mass becomes larger, the number of bound states with fixed quantum numbers in such a treatment increases; states that are close to the threshold for breakup into heavy mesons will emerge as eigenstates of this effective quantum mechanics problem.   These states all have wave functions that are dominantly in the long-range region.  Since these states have wave functions that are dominated by the known long-distance regime, one can deduce some of their properties--most critically their existence---even in the absence of detailed information about the effective potential at shorter distances.  

However,  this line of reasoning can break down: physically, these higher-lying states can decay into lower-lying tetraquarks via pion emission.  Thus the bound states of the effective quantum mechanical problem become resonant scattering states.      
 If the decay width of the state is broader than the energy level spacing between two neighboring tetraquark states, then the simple two-body quantum mechanical description is not valid.  However as shown in Ref.\cite{cai2019existence}, the width of these states is suppressed in the heavy quark limit.

The strategy is to at first assume the potential model description is valid and then show that the decay width of the state is parametrically narrower than the energy level spacing.  This is essentially the same logic as the treatment of the atomic spectrum of hydrogen:  one first makes a potential model description involving the electron and proton even though physically the hydrogen atom is coupled to the electromagnetic field allowing excited hydrogen states to decay via photon emission and subsequently uses Fermi's golden rule to show that the decay widths due to photon emission are small compared to level spacings. 

In Ref.\cite{cai2019existence}, only the long-distance interaction was considered in detail; the shorter-distance effective potential is well defined in the heavy quark limit and its properties, while affecting the details of high-lying states, including the precise tetraquark spectrum, does not affect many of their qualitative features (such as the level spacing or the existence of tetraquarks parametrically close to the threshold for the breakup into two heavy mesons).  

As we will see in the following section the role of the short-distance interaction becomes problematic for the $q\Bar{q}'Q\Bar{Q}$ case.   In that situation even for systems with exotic quantum numbers (assuming heavy quarks and antiquarks are separately conserved), a short-distance effective potential for a two-body effective quantum mechanical problem is not well defined in the same way as in the $\Bar{q}\Bar{q}'QQ$ case.  This problem is easy to evade for systems with sufficiently high orbital angular momenta; in those cases, the wave function becomes exponentially small prior to entering the regime where the effective potential becomes ill defined.  For the more general case, it is problematic.

Let us return to the case of tetraquarks with two heavy quarks.   Here we outline a few of the most salient features for that problem with an emphasis on those that are critical to the current problem.  We have altered the notation from Ref.\cite{cai2019existence} somewhat here to simplify the presentation; the underlying physics is the same.  While the formalism is easily adaptable to the case where the two heavy quarks have different masses, for simplicity the discussion below applies to the case where two heavy quarks are of the same type.  

In this discussion, the phrase ``fast degrees of freedom'' refers collectively to the light quarks and the gluons.

\begin{enumerate}

\item The treatment is based on a Hamiltonian formulation of QCD (with appropriate gauge fixing).  It is assumed formally that the theory is appropriately regulated in a gauge-invariant way ({\it e.g.} lattice regularized)  with the regulator at a scale of masses much larger than the mass of the heavy quark and appropriately renormalized.    To keep the analysis straightforward, we will take physical space to be a very large, but finite, physical box and will impose periodic boundary conditions on the box.  These periodic boundary conditions ensure that the total momentum remains a good quantum number, while the finiteness of the system forces it to discrete values.  

 In a Hamiltonian approach, the full Hilbert space is larger than the physical one since Gauss's law is imposed as a constraint.  In the treatment of exotic tetraquarks,  one only considers states in the physical Hilbert space.  Moreover, the entire treatment is restricted to a subspace of the physical space with the exotic quantum numbers of interest, {\it e.g.} doubly charmed states with isospin one; moreover, we restrict attention on states with total momentum zero, thereby decoupling center-of-mass motion from the internal motion of the state.
 
\item The analysis assumes that the heavy quark mass is large.   Pair creation of heavy quarks is suppressed in the limit where the quark masses become large for phenomena at a typical hadronic scale, $\Lambda$; the number of heavy quarks and antiquarks are separately conserved.  Thus if one fixes quantum numbers to states with charm or bottom of two, and takes the formal heavy quark limit, one knows that the state contains exactly two charm (or bottom)  quarks and no charm (or bottom) antiquarks.   

In principle at a large but finite heavy quark mass, one can integrate out the explicit heavy quark-antiquark degrees of freedom while accounting for their effects systematically in an expansion in $\frac{\Lambda}{m_H}$ corrections in an effective field theory treatment.   However, given the complexity of the problem and our qualitative goals, we worked to lowest order in such an expansion, neglecting all effects at relative order  $\frac{\Lambda}{m_H}$. 

Note that, at leading order  $\frac{\Lambda}{m_H}$, the spin of the heavy quark completely decouples\cite{IW} and plays no role in the analysis.   In particular, this means that states that differ only in the heavy quark spin quantum number---such as pseudoscalar and vector heavy mesons---will be treated as being degenerate.  Accordingly, at the order to which we are working, they may come in multiplates and in those cases one can use the heavy quark spin quantum numbers at the end of the problem to construct physical states with a variety of physical quantum numbers.

\item  The analysis depends on a systematic power counting scheme for describing dynamics at the hadronic scales,
 $\Lambda$.  The essential question involved how things scale as the heavy quark mass is made large.  For answering questions of this sort,  chiral scales of order $m_\pi$ that are insensitive to the heavy quark mass are simply taken to be of hadronic scale but numerically small.

 \item States in the physical Hilbert space with zero total momentum and the two heavy quarks at fixed distance $r$ from each other are denoted $|r, \psi \rangle$ where $\psi$ denotes all of the information about the fast degrees of freedom in the state.  A state is considered to be of this form if, and only if, it is in the physical subspace and 
\begin{equation}
\begin{split}
&\hat{R} |r ,\psi \rangle  =  r |r , \psi \rangle \; \; {\rm with} \\
& \hat{ R}    \equiv  \frac{ \int {\rm d}^3 x  \, {\rm d}^3 y  \, |y|  \,  \hat{Q}^\dagger(\vec{x}) \hat{Q}^\dagger(\vec{x}+\vec{y}) \hat{Q}(\vec{x}+\vec{y}) \hat{Q}(\vec{x}) }{2}  \; ,
\end{split} 
\label{Rdef}
\end{equation}
where $\hat{Q}$ is the nonrelativistic heavy quark field operator and $\hat{R}$ is the gauge-invariant operator that measures the distance between the heavy quarks.

 \item  The treatment was based on the following power counting, which was  assumed at the outset and subsequently shown to be self-consistent.
 \begin{equation}
 \begin{split}
 r &\sim \frac{1}{\Lambda} \\
 p_Q&\sim \sqrt{\Lambda m_Q} \\
 v_Q  &\sim \sqrt{\frac{\Lambda}{ m_Q}} ;
  \end{split}
\label{scaling}  \end{equation}
 $r$ is the separation of the heavy quarks,  $p_Q$ is the momentum conjugate to $r$ and $v_Q$ is the relative velocity.

This power counting is easily understood at an intuitive level  in term of the effective Born-Oppenheimer potential and semiclassical dynamics.  Such a potential is of characteristic depth $\Lambda$ and varies over a characteristic  length scale $1/\Lambda$, except at parametrically small distances.   If one studies a state with an energy of order $\Lambda$ below threshold, then a semiclassical analysis yields these scalings.  Although these scalings fail at small distances where the system is very near tuning points; the semiclassical wave function in both of these regions is parametrically small.

Note that this scaling behavior holds for states that are close to threshold in the sense that the binding energy is of order $m_\pi$, since as noted above,  the pion mass is simply treated as a lighter than usual hadronic scale. However it is does not hold for states that are extremely close to threshold in the parametric sense that the state is below threshold by an amount that goes to zero as $M_Q \rightarrow \infty$.  The parametrically near threshold case is discussed extensively in Ref.\cite{cai2019existence}, but is not relevant to the discussion in the present paper where this special case is not considered. 

 \item  It is assumed and subsequently shown to be self-consistent that a nonrelativistic potential model description of the dynamics is accurate up to corrections of relative order $\frac{\Lambda}{m_H}$.   The demonstration of self-consistency depends on the adiabatic motion of the heavy quarks and is based on the logic of the Born-Oppenheimer approximation \cite{born1927quantentheorie}.   The essential logic underlying this is that the scaling in \ref{scaling} implies that as $m_Q$ gets large the heavy quarks move very slowly allowing the fast quark and gluon degrees of freedom to continuously adjust so that they remain in their lowest-energy configurations.\label{selfcons}

\item Under the assumption that the potential model description is valid, one needs to further assume that the potential $V(r)$ depends on a single dynamical degree of freedom $r$, which is the distance between the two heavy mesons. Then if the $V(r)$ is attractive, and the possible bound state cannot decay to other states, the system must have a large number of bound states at large enough heavy quark mass. 

This argument can be shown based on simple treatment with the Schr\"odinger equation and variational method. The condition of a single dynamical degree of freedom is satisfied for systems with quantum numbers such as $I=1$, $J=0$, $S_Q=0$, and positive parity.

\item For systems with other quantum numbers, coupled channel effects might seem to render problematic the previous argument, which is based on a single slow degree of freedom. However, it can be shown that one can always find a local basis in which the channels decouple up to corrections that vanish in the heavy quark limit. Then multiple degrees of freedom reduce back to a single radial degree of freedom. Thus the previous argument still holds.

\item If there are open channels into which the bound state in previous arguments can decay, the previous potential model description breaks down. However, one can show that, for the case of $\Bar{q}\Bar{q}'QQ$, even though the state can decay to lower states by the emission of pions, the ratio of decay width to energy level spacing is parametrically suppressed by heavy quark mass. Thus in the heavy quark limit, there exist parametrically narrow tetraquarks of this type. Notice that, now one can say that in an appropriate regime the assumptions are indeed valid, and the logic of the argument is complete.
\end{enumerate}

Note that, although self-consistent assumptions of the validity of the potential model description are used, the argument does not depend on the details of any specific model. The theoretical framework can be built up from QCD. Here we give a sketch of some of the key elements of this framework that are particularly important for understanding the difference between the $\Bar{q}\Bar{q}'QQ$ and ${q}\Bar{q}' Q\Bar{Q}$ cases, further technical details can be found in Ref.\cite{cai2019existence}. 


The operator $\hat{R}$ measures the distance between the heavy quarks, so a state with $r$ fixed is an eigenstate of this operator in the whole Hilbert space.  Since $\hat{R}$ only tells us about heavy quarks, it gives no information about the state of the fast degrees of freedom (the light quarks and gluons)---$\psi_{\rm fast}$. Thus, we can denote eigenvector of $\hat{R}$ and total momentum operator $\hat{\vec{P}}$ with $\hat{R} |r, \psi_{\rm fast}; \vec{P}\rangle = r|r, \psi_{\rm fast};\vec{P}\rangle$.  For simplicity, the analysis will be done for a large spatial volume with periodic boundary conditions.  The periodic boundary ensures that the total momentum remains a good quantum number.  Without loss of generality, one can restrict one's attention to states with $\vec{P}=0$.  Accordingly, we will omit any reference to the total momentum of the state.

The next step is the construction of a projection operator $\hat{P}^r$ that projects the state onto a basis where the system is described by the Schr\"odinger equation with single variable $r$ as
\begin{align}
     &\hat{{\cal P}}^{\rm r} = \int _0^\infty {\rm d} r \,  |r, \psi_{\rm fast}^{\rm opt}(r) \rangle\langle r, \psi_{\rm fast}^{\rm opt}(r)\ |, 
\end{align}
where the label ``opt'' implies that this is the optimal state for the Schr\"odinger equation description.  To determine $\psi_{\rm fast}^{\rm opt}(r)$, one first constructs a potential operator by $\hat{V} \equiv \Hat{H} - \int d^3 x Q^\dagger(\vec{x})\left (-\frac{\vec{D}^2}{2 m_Q} \right) Q(\vec{x})$, where $\hat{H}$ is the NRQCD Hamiltonian. The potential function depends on the single kinetic degree of freedom $r$  and is given by:
\begin{equation}
V_{\psi_{\rm fast}} ( r)   \equiv \int d r' \langle r',\psi_{\rm fast}|\hat{V}| r, \psi_{\rm fast} \rangle  \; .
\end{equation}
Under the Born-Oppenheimer approximation, the optimal choice  $\psi_{\rm fast}^{\rm opt}(r)$ is the state that minimizes $V_{\psi_{\rm fast}} ( r)$.   Implicit in this, is the assumption that for fixed values of $\hat{R}$, the spectrum of the fast degrees of freedom has a gap in energy of order unity.  In that case the optimal choice is well defined and the adiabatic assumption underlying the Born-Oppenheimer approximation remains valid.  It is this assumption of a gap that breaks down in the case of $Q \overline{Q} q \overline{q}$ channels.

Using the projection operator, one can break the full physical Hamiltonian into the dominant term $\hat{H}_0$, which measures the energy of a heavy meson ``molecular'' states, and the interaction term $\hat{H}_I$, which induces the transition from higher-energy states to lower-energy levels. The key to the remaining analysis is the use of Fermi's golden rule and appropriate scaling rules to derive the scaling of decay rate $\Gamma$ based on $\hat{H}_I$'s effect on $| r, \psi_{\rm fast}\rangle$. 

The behavior of $\hat{H}_I | r, \psi_{\rm fast}\rangle$ can be deduced by noticing that $\hat{H}_I$ is obtained by subtracting the kinetic energy of heavy quarks $\hat{T}^{\text{heavy}}$ from the total kinetic energy of the tetraquark system, $-\frac{D^2_r}{M_Q}$, where $D_r$ is the covariant derivative acting on the single degree of freedom $r$. Using the known scaling rules, one can replace $-\frac{D^2_r}{M_Q}$ by $-\frac{\partial^2_r}{M_Q}$ since the difference is order $\mathcal{O}(\frac{1}{M_Q})$, where the derivative acting both explicitly on $r$ and implicitly on $\psi_{\rm fast}^{\rm opt}(r)$. 

Finally one can insert $\hat{H}_I | r, \psi_{\rm fast}\rangle$ into a formula derived for $\Gamma$ using Fermi's golden rule, and use semiclassical approximation form of the wave function to derive the scaling rule of decay rate as $\Gamma \sim \frac{\Lambda^2}{m_Q}$ when the binding energy $B$ is of order $\Lambda$. 
The derivation of this last scaling behavior is both long and intricate and will be omitted here; see Ref.~\cite{cai2019existence} for the details. This scaling does not hold for states that are parametrically close to threshold in the sense that they go to threshold as $M_Q \rightarrow \infty$  but does hold for states of order $m_\pi$ below threshold.
This scaling yields widths for the states that are parametrically small compared to the level spacing yielding clearly discernible resonances. Thus the logic of treating these resonant states as would-be bound states that decay slowly is self-consistent.



\section{logic of this work}

The arguments reviewed in the previous section apply to tetraquarks of type $\Bar{q}\Bar{q}'QQ$, while the focus of the current work is on  $q\Bar{q}'Q\Bar{Q}$  state. The latter case is problematic due to rearranging effects.  There are two kinds of final states into which $q\Bar{q}'Q\Bar{Q}$ tetraquark resonances can decay by strong interactions; one is a light meson (or mesons) plus more deeply bound tetraquarks, which is analogous to the decay of $\Bar{q}\Bar{q}'QQ$; the other is a light meson (or mesons) plus heavy quarkonium. For the first kind of decay channel, one can use semiclassical analysis and Born-Oppenheimer approximation and follow exactly the same logic as in Ref.\cite{cai2019existence} to show the decay width induced by this channel is parametrically suppressed by the heavy quark mass. However the second kind of decay channel, the heavy quarkonium is beyond the regime of bound state problem of two heavy mesons, and requires a more nuanced treatment.   The key issue is that the existence of the rearrangement channel implies that the assumption of a gapless spectrum is wrong.  This is the main focus of this paper.

To start with, we assume that the mass of the heavy quarks is sufficiently heavy that the dynamical creation of heavy quark-antiquark pairs is suppressed.  Thus the numbers of heavy quarks and heavy antiquarks are separately conserved.  Note that this condition should be true up to corrections of relative order $\Lambda/M_Q$.  Since the analysis of Ref.~\cite{cai2019existence} is only valid to relative order $\Lambda/M_Q$, the assumption that the numbers of heavy quarks and heavy antiquarks are separately conserved is valid to the order at which we are working.  Thus one can define a QCD operator  $\hat{ R}$  analogous to Eq.~(\ref{Rdef}) that gives the distance between the heavy quark and antiquark:
\begin{align}
     \hat{ R}   \equiv  \int {\rm d}^3 x  \, {\rm d}^3 y  \, |y|  \,  \hat{{Q}}(\vec{x}) \hat{\overline{Q}}^\dagger(\vec{x}+\vec{y}) \hat{\overline{Q}}(\vec{x}+\vec{y}) \hat{Q}(\vec{x})  \; ,\label{Rdef_new}
\end{align}
where $Q$ and $\overline{Q}$ are nonrelativistic two-component spinors.

We are considering the lowest-energy configuration of the system with charge quark contents $q\Bar{q}'Q\Bar{Q}$,  with $q$ and $q'$ distinct (so that the system has isospin 1).   Suppose that one could legitimately define an effective potential as a function of the distance between the heavy quarks $r$, where $r$ is an eigenvalue of $\hat{R}$.  At large $r$ the effective potential of the system would act completely analogously to the case of the $\Bar{q}\Bar{q}' {Q Q}$ channel: two heavy mesons $q\Bar{Q}$ and $\Bar{q}'Q$ that are interacting via a one-pion-exchange potential plus corrections that are exponentially small in the distance.  The spectrum for the fast degrees of freedom with fixed $r$ is gapped.  One needs enough energy to create an additional pion to create an excitation.   

In contrast, at small enough separation, the physical situation is radically different.  The physics of the lowest-energy configuration would be the heavy quark and antiquark forming a color singlet as in heavy quarkonium plus the addition of a pion that carries the isospin. Since the pion can carry a continuous range of energy, the energy spectrum of the system at the short-distance regime is gapless.  This means that there is no clean cut between fast and slow degrees of freedom, and the Born-Oppenheimer approximation is no longer valid. Thus, the spectrum of the light degrees of freedom at fixed $r$ is gapless when $r$ is small and becomes gapped when $r$ is large enough.  Clearly, there  is some value of $r$ that corresponds to the transition between the gapped and ungapped regimes

While this distinction between the long-distance gapped regime and the short-distance ungapped one, is based on a physical picture, the analysis that flows from it is nevertheless model independent.  We need not attempt to model either of these regimes except in a manner deducible from standard QCD results.

The difficulty in applying our analysis is that the Born-Oppenheimer approximation on which it is based depends on the spectrum of the fast degrees of freedom being gapped, while the physics indicates that it only gapped for sufficiently large $r$.  Moreover, typically one expects that the dynamics of the system when it is at large $r$, will push toward small $r$ since the potential function will be more attractive at small $r$ than at large $r$.  Thus, under typical circumstances one expects the analysis to be invalid and one cannot use it to predict the existence of long-lived tetraquark resonances.  

However, there is a regime where this typical expectation does not hold and one can use a very similar analysis to the  $\Bar{q}\Bar{q}' {Q Q}$.  The logic is similar to the demonstration in the previous section: one can at first assume the system is in an ``appropriate regime'', then show that within this regime the tetraquark of $q\Bar{q}'Q\Bar{Q}$ type exists, and finally shows the appropriate regime is indeed valid because the tetraquark is stable and remains in the same regime for long times.  The appropriate regime in  $q\Bar{q}'Q\Bar{Q}$ case is one in which $r$ is large enough so that the system is away from the gapless region and, significantly the system remains away from the gapless region. Then if one can show parametrically narrow $q\Bar{q}'Q\Bar{Q}$ state exists in this domain with large enough $r$ and does not decay to the gapless domain except over parametrically long times, the assumption is self-consistent.

 If the wave function stays away from the gapless domain, then as $m_Q$ goes to infinity, the problem is reduced to a single kinetic degree of freedom where corrections vanish in the infinite $m_Q$ limit\cite{cai2019existence}. The key point is that if the effective potential, $V(r)$ (which at large $m_Q$ is well defined when the light degrees of freedom are gapless as occurs at large $r$)  happens to have to a local minimum at some value $r_0$ within the gapped region, then as $M_Q \rightarrow \infty$ well-localized semiclassical states exist within this local minima.  Such semiclassical states are localized in the energetically allowed region within the well.  Of course, such states are not absolutely stable.  The system can tunnel out of such a state quantum mechanically.    However, the rate of such quantum tunneling becomes exponentially small as  $M_Q \rightarrow \infty$ leading to long-lived states.  Strictly, one can only conclude the decay rates are parametrically of order $\Lambda^2/M_Q$ since the entire formalism giving rise to the potential was only developed at leading order in $\Lambda/M_Q$.  Nevertheless this is enough to conclude that parametrically narrow tetraquarks exist.  Moreover, the conclusion that tetraquarks exist remains valid, even though the quantum mechanical tunneling leads to a regime in which the system can no longer be described by a single degree of freedom associated with distance.

Significantly, a regime of quantum numbers exists in which such a local minimum will necessarily form as $m_Q$ gets large provided that $m_\pi$ is sufficiently small.  This occurs in a regime when the angular momentum of the state $l$ is large, but not too large.  To understand intuitively why this is true, consider a three-dimensional quantum mechanical system of two particles interacting via a central potential.   The problem reduces to a radial problem with an  effective potential given by
\begin{equation}
V_{\rm eff}(r)=V(r) + \frac{l(l+1)}{2\mu r^2}
\end{equation}
where $\mu  \sim m_Q$ is the reduced mass and the second term, the centrifugal barrier, is repulsive.  Next, suppose that at the short-distance regime the magnitude of the potential grows with decreasing $r$ more slowly than $1/r^2$ and at the long-distance regime it is attractive and drops off faster than any power law; the Yukawa potential satisfies this.  The potential has the property that a repulsive centrifugal barrier will dominate $V_{eff}$ at both large $r$ and small $r$ provided that $L \ne 0$.  In some intermediate regions, the attractive bare potential may be able to overwhelm the centrifugal barrier leading to a local minimum in the effective potential.  The limited range in $L$ is easily understood.  On the one hand, if $L$ is too large, the centrifugal barrier will dominate over the bare potential everywhere and the local minimum is lost.  On the other, as $L$  decreases, the minimum of the effective potential will move inward since the centrifugal repulsion diminishes.  In the current context, it is important that the minimum of the effective potential is at a sufficiently long distance that the system is within the gapped region.

In the present problem, one can show self-consistently that there is a regime of angular momentum where a minimum in the effective potential exists at sufficiently long range as to be outside the gapless region, provided that the pion mass is sufficiently small.  The key point is that for sufficiently large $r$ the system becomes arbitrarily well described by two heavy mesons interacting via pion exchange; hence to good approximation $V(r)$ in this regime is well described by a Yukawa potential.  Suppose one were to neglect the short-distance non-Yukawa parts of the potential and find that for certain values of angular momentum (which can depend on $m_Q$), there is a minimum in the effective potential at large $m_Q$ that occurs at $r$ equals $c /m_\pi$ where $c$ is a numerical constant.  In that case, for sufficiently small $m_\pi$ the minimum would occur in a regime where the one-pion-exchange potential is indeed dominant and one has a self-consistent result.  Moreover, such a minimum is clearly outside the gapless region. If one can show that such a situation occurs, then one would have demonstrated the existence of parametrically narrow $q\Bar{q}'Q\Bar{Q}$  at large $m_Q$ provided that $m_{\pi}$ is sufficiently small.

 In the next section, it will be shown that this is precisely what happens.   There exist values of the angular momentum such that the minimum at $r$ is as large as $\frac{\phi}{m_{\pi}}$, where $\phi$ is  the golden ratio $\phi \approx 1.618$.  Formally this is sufficient to demonstrate the existence of a regime in which parametrically narrow $q\Bar{q}'Q\Bar{Q}$ exist.   Moreover, using the physical values of $m_\pi$, one has $\frac{\phi}{m_{\pi}} \approx 2.28 \, \text{fm}$.  Our phenomenological experience with the nucleon-nucleon potential\cite{ishii2007nuclear} strongly suggests that this is indeed large enough for the one-pion-exchange potential to dominate.   Thus, it is highly plausible that one only needs to consider the $m_Q \rightarrow \infty$ limit while keeping the light quark masses (and hence $m_\pi$) fixed.

As noted above, the angular momentum needs to be in a fixed range of angular momenta, neither too large nor too small.   As will be shown in the following section, the entire range of  values of $l$ for which this occurs is large at large $m_Q$:
 \begin{equation}
 l \sim \sqrt{\frac{m_Q}{\Lambda}}.
 \end{equation}

It is worth noting that, even if an exceptionally strong short-distance interaction were to cause the one-pion-exchange interaction with physical values of $m_\pi$ not to be dominant in the region of the minimum, the final argument about the existence of the tetraquark would continue to hold, provided the correction to the potential is small so that the exact location of the potential well remains inside the gapped regime.

\section{existence of a local minimum in the effective potential}

The purpose of this section is to show that, given the two assumptions in the previous section, in the heavy quark mass limit, a local minimum in the effective potential must exist for certain channels within a range of large angular momenta and the decay chance from bound states in this potential well to gapless spectrum domain is parametrically small; this implies that the assumptions are self-consistent. 


When $r$ is sufficiently large, the system becomes increasingly well described by two heavy mesons interacting via a Yukawa-like one-pion-exchange potential. This follows trivially from the fact that the pion is the lightest degree of freedom in QCD.  Thus we begin by ignoring the short-distance interaction and will show {\it a posteriori} that this is consistent.  There are a number of spin and isospin configurations possible and we will choose quantum numbers for which the one-pion-exchange potential is attractive.   In order to show that a minimum in the effective potential exists for certain angular momenta, one needs to notice that, when angular momentum $L$ is of order $(m_Q/\Lambda)^{1/2}$, the centrifugal term $\frac{1}{m_Q}\frac{l(l+1)}{r^2}$ plays a significant dynamical role. This term has the opposite sign from the Yukawa potential and dominates at sufficiently large distances. 

One might worry that the tensor force may play a role so that there are multiple degrees of freedom.  However, using the same argument as in Ref.\cite{cai2019existence},  one can show that, it can be reduced to one degree of freedom with corrections that vanish in the heavy quark mass limit. A key point in generalizing that analysis to the current case is that, we are looking for a regime, for which $m_Q$ is large, and $L$ grows like $\sqrt{\frac{m_Q}{\Lambda}}$, while other quantum numbers are still of order $1$.  The tensor force mixes $l$s that differ by 2.  Thus the value of the centrifugal barrier in the terms that mixed by the tensor force is the same up to subleading corrections in $\sqrt{\frac{\Lambda}{m_Q}}$.

In this case, the effective potential is described by
\begin{align}\label{potential1}
    V_{\rm eff}(r) = - a\frac{e^{-m_{\pi}r}}{r} +  \frac{l(l+1)}{m_H r^2},
\end{align}
where $a$ is some constant that depends on the coupling of pions to the heavy mesons, and $m_H$ is the mass of the heavy meson. $m_H$ is equal to the heavy quark mass $m_Q$ with corrections that vanish in heavy quark limit.

At extremely large $r$, the first term of the effective potential is negligible and the magnitude decreases much faster than the first term; while when $r$ is not extremely large, the first term could be comparable to the size of the second term for a certain range of $r$, and the second term may decrease faster than the first term. Thus, one can imagine if a local minimum exists, it would be at a relatively large $r$, so that the Yukawa term is reasonable, but not extremely large $r$. 

As a formal matter, once there is a local minimum in the effective potential, no matter how shallow the local potential well,  as the $m_Q \rightarrow \infty$ limit is approached, the system will have an arbitrarily large number of would-be bound states that become increasingly narrow as the limit is approached.  This follows from a trivial semiclassical argument. 

Let us define  $r_{\text{loc min}}$ as giving the position of the local minimum  and $r_{\text{sep}}$ is the value of $r$ that separates the gapless region from the gapped region.   The value of $r_{\text{sep}}$ is given  implicitly by
\begin{equation}
V^{I=1}_{\text{Bo-Opp}}(r_{\text{sep}} )=V^{I=0}_{\text{Bo-Opp}}(r_{\text{sep}} )+ m_{\pi}
\end{equation}
where $V^{I=1}_{\text{Bo-Opp}}$ and  $V^{I=0}_{\text{Bo-Opp}}$ are the Born-Oppenheimer potentials computed for the isovector and isoscalar channels respectively.

 If the $r_{\text{loc min}}$ is notably larger than $r_{\text{sep}}$, the tunneling from the two heavy meson state at $r_{\text{loc min}}$ to heavy quarkonium plus pion at $r_{\text{sep}}$ is exponentially suppressed. As noted earlier,  there is a subtle caveat:  the previous arguments are in the context of ignoring $\frac{1}{m_Q}$ correction. Thus, to the order at which we are working, the actual decay rate is only known to be suppressed by  $\frac{1}{m_Q}$, but this is sufficient to demonstrate the existence of parametrically narrow tetraquark states. 

Clearly, the position of $r_{\text{loc min}}$ is critical; we wish to ensure that $r_{\text{loc min}}$ to be as large as possible in order to ensure that it is both cleanly separated safely from the gapless region ({\it i.e.} that  $r_{\text{loc min}} > r_{\text{sep}}$) and inside the regime dominated by the one-pion-exchange potential to ensure self-consistency. The question now becomes finding the minimum point $r_{\text{loc min}}(a, l)$ of Eq.~\ref{potential1}, and tuning $l$ to obtain the maximum value of $r_{\text{loc min}}(a, l)$. 

In doing this, we will treat $l$ as if it is a continuous variable.  One may worry that in reality, $l$ takes discrete values, and thus the maximum value of  $r_{\text{loc min}}$  obtained by taking $l$ continuous might be larger than the true value. However,  in the heavy quark mass limit, the coefficient combination $\frac{l(l+1)}{m_Q}$ can be tuned to arbitrarily close to any positive value, so that it is effectively continuous; any error we make in doing this will be of order $\Lambda/m_Q$ and we have neglected effects of this order throughout the analysis.

The simplest way  to determine the $l$ leading to the largest value of   $r_{\text{loc min}}$,  is to start by  expressing all distances in terms of a dimensionless variable $t \equiv m_\pi r$:
\begin{align}\label{dimless}
    f(t) \equiv \frac{V(\frac{t}{m_{\pi}})}{a m_{\pi}} = - \frac{e^{-t}}{t} + \frac{k}{t^2},  
\end{align}
where  $k\equiv \frac{l(l+1) m_{\pi}}{2a m_H}$. When $m_H$ is large enough, by varying $l$, $k$ can be arbitrarily close to any positive value.  Note that when $k$ is a number of order unity, $l \sim  \sqrt{\frac{m_Q}{\Lambda}}$.   The position of the local minimum, if one exists, is the value of $t$ at which $f'(t)=0$.

Denote $t_{\text{loc min}}(k) \equiv m_{\pi}r_{\text{loc min}}$ as the position of minimum point of $f(t)$, and denote the largest value of $t_{\text{loc min}}(k)$ as $t_{\text{loc min}}^{\text{max}}$,
\begin{align}
    t_{\text{loc min}}^{\text{max}} \equiv \max_{k\in(0, \infty)}\{ t_{\text{loc min}}(k) \}= \max_{k\in(0, \infty)}\{ \operatorname*{arg\,min}_{t\in (0, \infty)} f(t) \}
\end{align}

After some simple algebra, one finds $t_{\text{loc min}}^{\text{max}} = \phi\equiv \frac{1+\sqrt{5}}{2}$.  Remarkably this is the golden ratio.  For the physical value of $m_\pi$ this  corresponds to  $r_{\text{loc min}}^{\text{max}} \equiv \frac{t_{\text{loc min}}^{\text{max}}}{m_{\pi}}\approx 2.28 \, \text{fm}$.    At such $r_{\text{loc min}}^{\text{max}} > 2 \, \text{fm}$.  Thus,  the previous assumption that the one-pion-exchange Yukawa potential should be dominant over the short-distance interaction is valid, given our experiences with nucleon-nucleon potential. Of course, one can imagine models where the short-distance interaction has an exceptionally large value even as far out as 2 fm.  However, even in that case there is still a formal regime in which the one-pion exchange dominates:  if one considers a version of QCD in which the light quark masses can be adjusted, then by making $m_\pi$ sufficiently light, one  pushes the putative  minimum associated with one-pion-exchange further out, and, since the one-pion exchange potential drops off with distance more slowly than the short-distance potential, one eventually gets to a region in which the one-pion exchange dominates.

When $m_H$ is large, there would be a large number of heavy tetraquark levels around $t_{\text{loc min}}^{\text{max}}$, assuming decays are suppressed. Note there are two kinds of decays, one is decay to a lower-level tetraquark state by emitting a light meson or multiple light mesons, and the other is decay to a heavy quarkonium state by emitting a light meson or mesons.

If one ignores the second case, the ratio of decay rate to energy level spacing of the first case is already shown to be parametrically suppressed by $m_H$ in Ref.\cite{cai2019existence}, where the charge quark configuration is different, but the proof is exactly the same. 


For the second case, it is reasonable to assume $r_{\text{sep}} < r_{\text{loc min}}^{\text{max}} $ and $r_{\text{sep}} \sim \frac{1}{\Lambda}$, where $\Lambda$ is the typical hadronic scale. The tetraquark state wave function  centered at $r_{\text{optimal}} \approx  2.28 \, \text{fm}$, is of order $\frac{1}{\Lambda}$ separated from the heavy quarkonium plus light meson(s) wave function below $r_{\text{sep}}$. Based on semiclassical analysis, it is obvious that the tunneling rate is exponentially small in $m_H \approx m_Q$, provided that one can treat the problem as a simple quantum mechanical problem with a single  relevant degree of freedom. A simple way to see this is that if one were to arbitrarily set the wave function of the two-heavy-meson state at $r_{\text{sep}}$ to zero as  a boundary condition, then the overlapping of the wave function of the two-heavy-meson state and the wave function of the heavy quarkonium plus light meson(s) state would be zero. The error caused by setting the wave function at the boundary to be 0 is exponentially small because $r_{\text{sep}}$ is at the classical forbidden area.

Since the decay to both of these two channels are suppressed by $m_Q$ in the heavy quark limit, the tetraquark resonance is parametrically stable at the heavy quark limit, so that the previous assumption that the state is in the domain where the spectrum  of the fast degree of freedom is not gapless is satisfied--at least if $m_\pi$ is sufficiently light. Now the proof is self-consistent, and the logic is complete.

It is worth noting the existence of an effective potential with a local minimum in the gapless region is, in effect a guarantee that for sufficiently large heavy quark masses, the heavy tetraquark can be described in the language of a model in which the dominant dynamics is the motion of the heavy quarks against each other with the heavy quarks dragging the light degree of freedom along.  Both molecular-type and diquark-type models are of this sort.  The formalism here cannot distinguish between these two (or indeed a description that interpolates between them) except that a molecular description becomes more plausible if the minimum of the potential well is both shallow and at large distances.

\section{discussion}

As stated in the introduction, we are primarily interested in exploring the conjecture that the existence of experimentally observed $q\Bar{q}'Q\Bar{Q}$ tetraquark states is tied directly to the large mass of the heavy quarks in them.   While in the previous sections, it is shown  in a model-independent way that  parametrically narrow $q\Bar{q}'Q\Bar{Q}$ tetraquark states exist as $m_Q$ gets large, these states only were shown to exist for a limited range of angular momenta, all of which were of order $\sqrt{m_q/\Lambda}$, and hence parametrically large.   Unfortunately, it is not straightforward to extend the argument to  the case of angular momenta of order unity---the case of phenomenological interest.   




To understand the challenge, recall that because the  heavy quark mass is formally taken to be arbitrarily large in the analysis, the fast (light quark and gluon) degrees of freedom and  the slow degree of freedom (associated with the separation of the heavy quark and antiquark)  have distinct timescales.  Thus one expects the adiabatic approximation to be  valid.  The slow degrees of freedom is sufficiently slow that the fast degrees of freedom have ample time to rearrange themselves into the lowest state in their spectrum---provided that they started in the lowest state and the spectrum is gapped  with the ground state well  separated from its first excitation.  Clearly, this last condition breaks down as soon as $r < r_{\text{sep}}$ and the spectrum becomes gapless.  This means that the analysis of the previous section breaks down unless the dominant region for the resonant behavior is separated from the gapless region.  Moreover as will be discussed below, not only does the analysis leading to the prediction of a parametrically narrow resonance  break down, the same reasoning implies that a resonance of this type does not exist. 

To see why, start by noting  that even in the gapless region the scaling rules of Eq.~(\ref{scaling}) still hold.  Thus, in the heavy quark limit, the slow degree of freedom remains parametrically slow.  While one can no longer conclude that the system will remain in its lowest state (since it is no longer isolated) one can conclude  that in the absence of any unexpected behavior  the scaling rules should ensure that the fast degree of freedom will remain in the low-lying part of its spectrum.  Moreover as $m_Q \rightarrow \infty$, one expects the populated part of this spectrum to become increasingly pushed toward its minimum.  Physically, all of these low-lying excited states correspond to the emission of at least one pion that carries away the isospin.   (Recall that our analysis focuses on states with isospin 1.)  Since the motion of the heavy degree of freedom is parametrically slow, one expects that the created pion will propagate a long distance away from the heavy quarks  before the slow degree of freedom has appreciably changed.  At such long distances, one expects the interaction between the emitted pion and the residual heavy quarkonium state to be negligible.
  
Thus when the mass is sufficiently large, the time evolution of a quantum state that initially has nontrivial weight in the region with $r<r_{\text{sep}}$ will involve the emission of a pion (or multiple pions) with very high probability on a timescale much faster than the scale by which the heavy degrees of freedom change appreciably (unless the state already has a pion at long distance).   If instead one starts with a state that has little overlap with the $r<r_{\text{sep}}$ one expects that the heavy degree of freedom in time evolving will head toward the region $r<r_{\text{sep}}$ since it has lower potential energy, at which point a pion is rapidly emitted. In  either case one does not expect such a state to be a useful model of a narrow tetraquark state.   The central underlying point is that when the quark mass is sufficiently large, and hence, its motion sufficiently slow, there will be enough time to emit a pion whenever the separation of the heavy mesons is in the regime where it is energetically favorable.

To go further, it is worth discussing in a qualitative way the nature of narrow resonant states in quantum mechanics and quantum field theory.  There are many ways to think about resonances.  For example, one can identify resonances as poles on the second sheet of the S-matrix analytically continued to an unphysical Riemann sheet; narrow resonances have poles with small imaginary parts.   However, for the present purposes, it is useful to focus on the underlying physical principles that allow narrow resonances to occur.  Crudely, they occur when a ``would-be'' bound state is weakly coupled to at least one energetically open channel.  Such a would-be bound state might arise in two-body scattering but can also occur in much more complicated situations.  This weak coupling can occur for many reasons.  One natural way is when the would-be bound state needs to tunnel through a potential barrier to reach the regime in which the open channel is accessible.  This  typically occurs for narrow resonances in two-body scattering problems (or equivalently one particle in a potential well)  in nonrelativistic quantum mechanics.  Such a situation can lead to resonances that are parametrically exponentially narrow as either the height of the barrier or its width gets large.  In three-dimensional problems with spherical symmetry,  the barrier in two-body scattering problems that leads to a narrow resonance often arises due to the centrifugal barrier in the effective potential.  An analogous situation happened in the present context in cases where the angular momentum was large.

There is a simple way to understand the behavior when the resonance is associated with the motion of a single degree of freedom in which a local minimum is bounded by two turning points, $r_{\text{min}}$ (which may be zero) and $r_{\text{max}}$, and where the mass is sufficiently heavy that the system can be thought of as being nonrelativistic and in the semiclassical limit.   At the classical level, the system oscillates with a frequency given 
\begin{equation}
\omega=\frac{2 \pi  \sqrt{\mu}} {2 \int_{r_{\text{min}}}^{r_{\text{max}}} dx \sqrt{2 (E-V^{\rm eff}(r)}}
\end{equation}
which at the semiclassical level gives  the splitting between different resonant states.  As the system approaches the turning points one can use standard semiclassical methods to estimate the probability that it will tunnel through the barrier and leave the resonance region; these probabilities are characteristically exponentially small.  If the probability that the system tunnels out of the region during one cycle is denoted $P_{\rm tun}$ and is much less than unity, then the decay rate of the resonant state is given by
\begin{equation}
\Gamma_{\rm decay}  =2 \pi  \omega P_{\rm tun}
\end{equation}
The condition for a resonance to be considered narrow is that
\begin{equation}
\begin{split}
Q & \gg 1  \; \; {\rm where} \\
Q & \equiv \frac {\omega}{\Gamma_{\rm decay}} = \frac{1}{2 \pi  P_{\rm tun}} \, ;
\end{split}
\end{equation}
$Q$ is defined by analogy to the Q factor in resonant cavities or RLC circuits.  Since the decay rate directly gives the width of the resonance, large $Q$ implies that the width of the resonance is small compared to the spacing.  

Consider the situation of a would-be bound state weakly coupling to an energetically open channel and focus on what happens when the coupling to the open channel increases.  Generically, the width of the resonance will increase.  In the case of nonrelativistic scattering of a particle in a potential well where the weak coupling is due to a potential barrier, if the height of the potential separating the would-be bound state from the open channel is decreased the resonance will broaden.  If that barrier is lowered to the point that it is completely gone, the would-be bound state will have dissolved into the open channel and the resonance lost.  

Let us now return to the problem at hand, putative heavy tetraquarks of the sort that emerge due to the Born-Oppenheimer approximation.  The would-be bound state needs to be localized in the gapless region where pion emission is suppressed.  There are two ways that  a narrow tetraquark resonance of this sort could be lost.   One way is that  the barrier separating the bottom of the well  of the effective potential with a configuration with two arbitrarily well-separated heavy quarks (corresponding to  widely separated mesons) could vanish.  The other is that the barrier that separates the minimum from the gapless region can vanish.  In either case, a narrow resonance would disappear provided that the quark mass is heavy enough for this type of analysis to hold.

In Fig.\ref{cartoon} this is illustrated schematically.  In the figure a schematic  Born-Oppenheimer effective potential is shown for various values of the angular momentum; all of the angular momenta are assumed to be of order $\sqrt{m_Q/\lambda}$.  Longer dashes on the lines correspond to higher angular momenta and are seen to be at higher values for fixed $r$.  The potential is not realistic and is given only for the purposes of illustrating qualitative features.  An important assumption regarding the  schematic potential  is that for zero angular momentum it is  monotonically increasing with increasing $r$, and eventually asymptotes to a fixed value.  Any potential of this sort  will yield qualitatively similar behavior.  The gray shaded region in the figure corresponds to the gapless domain where it is energetically favorable to emit a pion; it extends out to $r_{\text{sep}}$.  The key features seen in the figure are that for small values of the angular momentum, there is no local minimum in the gapped region. Beyond some lower critical value of the angular momentum--in the figure this corresponds to the second curve from the bottom---a local minimum forms (and with it a tetraquark resonance).    As the angular momentum increases from there,  the position of the effective potential pushes outward.  For comparatively small angular momenta beyond the lower critical value, the effective potential  at the local  minimum is below the value of the effective potential at infinite $r$--which corresponds to well-separated heavy mesons and is represented by the horizontal line in Fig.~\ref{cartoon}.   In these cases when the heavy quark mass is very large and there are resonant states with energies close to the value of the potential at the local minimum, the only open decay channel is via pion emission.   As the angular momentum increases  further, the centrifugal barrier  eventually causes the local minimum to occurs with a value above that of the effective potential at infinity.  A resonance of a heavy quark system that is associated with such a minimum can decay either via pion emission or by breakup into two heavy mesons.  Eventually, as the angular momentum increases still further, an upper critical value is reached in which the local minimum vanishes and with it, the tetraquark resonance.

As noted above the behavior in Fig.\ref{cartoon} is generic provided that the Born-Oppenheimer potential for zero angular momentum has the property that it is monotonically increasing with $r$ in the gapped region when the angular momentum is zero.  Clearly, when the angular momentum is beyond the upper critical value the resonance is simply lost.  The behavior when angular momentum is below the lower critical value requires a bit more thought.  In that case, the Born-Oppenheimer formalism by which the problem reduces to an effective problem with a single degree of freedom has broken.



In the case of a putative tetraquark state with small angular momentum (of order unity), the analysis for the case of a  $q\Bar{q}'Q\Bar{Q}$ tetraquark state clearly breaks down.
This does not necessarily mean that the conjecture that the experimentally-observed narrow near-threshold $q\Bar{q}'Q\Bar{Q}$ tetraquarks exist because of the large mass of the heavy quarks needs to be abandoned.   However, it is difficult to imagine how the conjecture is to hold unless a particular condition is met.  
 
 The scaling rules of Eq.~(\ref{scaling}) will continue to hold regardless of whether the state is resonant and narrow or not.  Thus, the general picture in which the slow motion of the heavy quarks with the light degrees of freedom rearranging based on the position of the heavy quarks will dominate the dynamics, again regardless of whether the state is resonant and narrow or not.  If there is no impediment to the system moving from the gapped to ungapped regions, the reason for a resonance of the sort discussed here vanishes. 
 
 One might consider instead scenarios in which the dynamics of the light degrees of freedom themselves are somehow responsible for the resonance.  Such scenarios have a certain initial plausibility.    For example, one might envision that the reason for the existence of the resonances lies in the very small overlap of states in which the light degrees of freedom form a pion and those in which the light quarks attach themselves to the heavy quarks forming heavy mesons.  It is conceivable that such an explanation may explain, at least partially, some or all of the tetraquark resonances observed experimentally.  However, such a mechanism is very unlikely to yield parametrically narrow resonances at large $m_Q$. 
 
  As a general matter, it is hard to see why any mechanism involving light degrees of freedom which would make resonances increasingly narrow at large $m_Q$ should exist.   Note that as the resonances get narrow the timescale gets longer; thus if the resonance is parametrically narrow at large $m_Q$, it must know about the heavy quark dynamics, but if the heavy quarks are involved dynamically, the motion will be adiabatic and one expects the picture developed here to hold.
 
 For the particular scenario where the small overlap of configurations in which the light degrees of freedom form a pion and those in which the light quarks are in heavy mesons, the situation is worse---as the heavy quark masses increase, such a mechanism ceases to work at all.  The reason for this is the adiabatic nature of the motion from regions in which the heavy quarks are well-separated  (where the system acts very much like two heavy mesons) to just before the beginning of the ungapped region (where the system acts very much like a pion very weakly bound to the heavy quark system) when the quarks are heavy.  While the configurations of the light degrees of freedom in these two regions presumably have an extremely small overlap, the system smoothly evolves from one of these to another; the overlap between them plays no role in the dynamics.  The physical reason for this is that when the masses are sufficiently large the heavy quarks are moving so slowly that the light degrees of freedom always have time to adjust locally to something quite close to the optimal configuration.  This allows the system to smoothly evolve from one qualitative type of light quark configuration to another without the small overlap between them acting as an impediment.

While it is not logically impossible that the dynamics of the light degrees of freedom could give rise to parametrically narrow resonances at large $m_Q$, it seems quite implausible.  If one eliminates such scenarios from consideration, there appear to be two classes of scenarios that might be viable for preserving the conjecture of the existence and parametric narrowness of low-angular-momentum $q\Bar{q}'Q\Bar{Q}$  tetraquark resonance at large values of $m_Q$.    The most natural type of scenario is that something very similar to what happens in the large angular momentum case would have to occur: the system's effective potential in the gapped region has a local minimum.  If this occurs, then the wave function of the system is trapped in the potential well; it becomes exponentially hard to tunnel to the short-distance region and the resonance becomes narrow, at least as $m_Q$ gets large.  Such a minimum would have to occur in a region shorter than the one-pion range, which has no such minimum at small angular momentum, but also must be outside of the gapped regime where the notion of an effective potential is sensible.   Of course, in this case,  unlike in the case of large angular momenta, the local minimum would need to arise due to some unknown dynamical reason.  While such a mechanism is unknown, there is no reason {\it a priori} that it could not occur.

However, if no such local minimum exists in the adiabatic effective potential, the nature of the narrow low-angular-momentum resonant states observed experimentally must be rather different from the type described here.   In particular, they cannot be due to adiabatic motion of the heavy quarks and necessarily require subleading effects in $1/m_Q$.  One ironic possibility is that these resonances simply disappear if the quark masses become too large; in that case, it is not the heaviness of these quarks that causes these resonances but rather their lightness.   Another possibility is that the resonances are somehow due to the dynamics of the light degrees of freedom.  However, as noted above, this possibility seems unlikely.

Fortunately, lattice studies can help answer the question of whether a local minimum in the adiabatic potential occurs.  The adiabatic potential for an $I=1$ system with fixed color sources can be computed on the lattice.  If sufficiently reliable lattice calculations are done, this should allow for a determination of whether such a local minimum occurs.  Of course, like all lattice studies, they are subject to statistical uncertainties as well as systematic ones associated with extrapolations to the infinite volume and continuum limits as possible extrapolations to the physical quark mass.  Preliminary lattice calculations of this adiabatic potential have been done\cite{Prelovsek:2019ywc}; they show no indication of a local minimum forming in the ungapped region.  If this is confirmed by more extensive calculations with all of the uncertainties under control, then this possibility can be largely eliminated.  Clearly, additional lattice studies are needed to clarify the situation.

If future lattice calculations of the adiabatic potential confirm this feature of the preliminary ones,  then one is forced to conclude that  $q\Bar{q}'Q\Bar{Q}$ systems with low-angular-momenta either have no parametrically narrow low-angular momentum tetraquarks in the heavy quark limit or, they exist but arise by some mechanism involving the light degrees of freedom that is unknown, and, given what we know, implausible.  If we exclude such implausible scenarios, we are led to consider a remarkable possibility: the existence of narrow  $q\Bar{q}'Q\Bar{Q}$ tetraquark resonances with low angular momenta, may due to the heavy quarks being sufficiently light;  rather than the resonances arising due to the extremely large value of $m_Q$ in the physical world, they arise precisely because these masses are small enough for their motion not to be adiabatic.  This might be the reason of why the bottom partner of $X(3872)$ is still unseen experimentally.

\acknowledgements

The authors thank the U.S. Department of Energy for supporting this research under Contract No. DE-FG02-93ER-40762.

\newpage

\bibliography{main}

\end{document}